\documentclass{article}
\usepackage[utf8]{inputenc}
\usepackage{times}
\usepackage{amsfonts}
\usepackage{amsmath,amssymb}
\usepackage{indentfirst}
\usepackage{wrapfig}
\usepackage{graphicx}
\usepackage{subcaption}
\usepackage{graphics,color}
\usepackage{caption}
\usepackage[toc,page]{appendix}
\usepackage{color}
\usepackage{cite}
\usepackage[inactive]{srcltx}
\usepackage[T1]{fontenc}
\usepackage{float}
\usepackage{hyperref}
\usepackage{manyfoot}%
\usepackage{booktabs}%
\usepackage{algorithm}%
\usepackage{algorithmicx}%
\usepackage{algpseudocode}%
\usepackage{listings}%
\usepackage{dsfont}%
\usepackage{ulem}%

\begin{document}
\date{}
\begin{center}
{\Large\bf Fock-state filtering of squeezed states: enhanced nonclassicality and effects of a thermal bath}
\end{center}
\begin{center}
{\normalsize João Pedro Gonzalez de Oliveira and A. Vidiella-Barranco \footnote{vidiella@ifi.unicamp.br}}
\end{center}
\begin{center}
	{\normalsize{$^{b}$Gleb Wataghin Institute of Physics - University of Campinas}}\\
	{\normalsize{ 13083-859   Campinas,  SP,  Brazil}}\\
\end{center}
\begin{abstract}
The ability to tailor photon-number statistics is central to the generation and control of nonclassical states of light. Here, we investigate how Fock-state filtering of squeezed coherent states can be used to engineer their photon statistics. We consider the selective removal of the $m=0$ and $m=1$ Fock components and show that this process can enhance the sub-Poissonian character of the field, depending on the coherent amplitude. We then analyze the evolution of these engineered states in a thermal environment. Although all states eventually relax toward thermal statistics, the signatures of the filtering process may persist over a significant part of the evolution, while the purity of the filtered states is rapidly degraded by incoherent population redistribution. Our results illustrate both the potential and the limitations of photon-number engineering in realistic, dissipative settings.
\end{abstract}
\section{Introduction}\label{sec1}

In 2025, the International Year of Quantum (IYQ), we celebrated not only 100 years of Quantum Mechanics, but also 45 years of the Caldeira-Leggett model, a paradigmatic model of quantum dissipation \cite{caldeiraleggett1981,caldeiraleggett1983a,caldeiraleggett1983b}, developed during A.O. Caldeira's Ph.D. under the supervision of A.J. Leggett, in the early 1980s. Since its inception, quantum theory has undergone extraordinary development, and in recent years it has advanced into areas related to technology. System–environment interactions are now recognized as playing a fundamental role in shaping both the properties and evolution of quantum systems, with the Caldeira-Leggett model providing one of the most general microscopic treatments of the interaction of quantum systems with a thermal bath modeled by a collection of harmonic oscillators. A standard approach in the theory of open quantum systems consists of deriving a {\it master equation} for the system's reduced density operator, $\rho_s(t)$. This is typically accomplished by starting from the von Neumann equation for the total system-bath density operator and tracing out the degrees of freedom of the bath. The resulting evolution equation, in the Schrödinger picture, can be written in the generic form
\begin{equation}
	\frac{d\rho_s(t)}{dt} = -\frac{i}{\hbar}[\hat{H},\rho_s(t)]+\hat{\cal L}\rho_s(t),\label{mastereqgen}
\end{equation}
where $\hat{H}$ is the system's Hamiltonian and $\hat{\cal L}$ a superoperator describing the influence of the environment. In contrast to the unitary evolution of the closed system, the reduced dynamics governed by Eq.(\ref{mastereqgen}) normally exhibits dissipation and decoherence. A variety of quantum master equations has been developed, each incorporating specific physical assumptions and approximations regarding the system–environment coupling and the properties of the bath \cite{breuer02}.

Quantum optics was one of the fields that helped to shape the modern theory of open quantum systems. One of the simplest examples is a single quantized cavity mode confined in a cavity, described by creation (annihilation) operator $\hat{a}^\dagger(\hat{a})$ obeying $[\hat{a},\hat{a}^\dagger] = \mathds{1}$. In this case, the Hamiltonian of the single-mode cavity of frequency $\omega$ in Eq.(\ref{mastereqgen}) is simply $ \hat{H} = \hbar \omega \hat{a}^\dagger \hat{a}$. Considering a non-ideal cavity (imperfect mirrors), the photon leakage into the exterior leads, upon tracing out the bath modes, to a master equation for the reduced density operator of the cavity field. 

Although the normally adopted master equations are less general than the Caldeira-Leggett master equation, they account for effects such as dissipation and decoherence in certain regimes. The quantization of light is not only intertwined with the beginnings of quantum theory, but also plays a fundamental role in so-called quantum technologies \cite{barnett2017}. Quantum optics offers a uniquely controlled framework in which quantum-mechanical phenomena, such as coherence and entanglement, can be investigated. While classical electrodynamics associates a single configuration to each cavity mode, in quantum electrodynamics that same mode spans an infinite-dimensional state vector space, allowing the preparation of a wide variety of distinct quantum states of light. We can define a few important quantum states of light (or equivalently, of a harmonic oscillator) associated with that single mode. A comprehensive account of the variety of quantum states of light can be found in Ref. \cite{dodonov02}. The eigenstates of the number operator $\hat{n} = \hat{a}^\dagger\hat{a}$, i.e., $\hat{n}|N\rangle = N |N\rangle$, are named Fock (or number) states $|N\rangle$. The Fock state $|N\rangle$ has exactly $N$ energy excitations (photons), where $N = 0, 1, 2,\ldots$ is a discrete parameter. These states are orthogonal, $\langle n|m\rangle = \delta_{n,m}$, and obey $\sum_{n=0}^\infty |n\rangle\langle n| = \mathds{1}$, forming a complete basis in Hilbert space. Fock states are also regarded as highly nonclassical states, given that the concept of discrete field quanta (photons) has no counterpart in classical electrodynamics. All the other quantum states of light exhibit some fluctuation in photon number. The coherent states are a particularly important class of states,  regarded as {\it quasi-classical} states of light. They possess a nonvanishing expectation value of the electric field operator whose time evolution coincides with that of a classical electromagnetic wave, although the field itself exhibits intrinsic quantum fluctuations. This contrasts with the Fock states, which have a zero expectation value for the electric field for any $N$. The coherent states $|\alpha\rangle$ can be defined as eigenstates of the annihilation operator $\hat{a}$, satisfying $\hat{a} |\alpha\rangle = \alpha |\alpha\rangle$, where $\alpha = |\alpha| e^{i\phi}$ is a complex parameter. It exhibits photon number fluctuations, with mean photon number $\langle \alpha | \hat{n} |\alpha\rangle = |\alpha|^2$ and variance $(\Delta n)^2 = |\alpha|^2$, i.e., the coherent states have a Poissonian photon number distribution. Another remarkable property of the coherent states is that the quadrature fluctuations are the same; the variances of the quadrature operators $\hat{X} = (\hat{a}^\dagger + \hat{a})/2$ and $\hat{X} = (\hat{a} - \hat{a}^\dagger)/2i$ are equal, which characterizes {\it minimum uncertainty states} in these variables. This is because the commutator $[\hat{X},\hat{Y}] = i/2$ implies $\Delta X \Delta Y \geq 1/4$, and the equality $\Delta X \Delta Y = 1/4$ represents the minimum uncertainty allowed within the quantum formalism (Heisenberg bound). Interestingly, the coherent states of the mechanical harmonic oscillator were introduced by E. Schrödinger in 1926 \cite{schroedinger1926}, and reintroduced in the quantum optical context by R.J. Glauber \cite{glauber1963} and E.C.G. Sudarshan \cite{sudarshan1963} in 1963. In fact, the first successful operation of a laser, achieved by T. H. Maiman in 1960, \cite{maiman1960} stimulated developments in the theory of coherent states of light, with the quantum state of laser radiation more accurately described as a phase-diffused coherent state \cite{scullyzubairy1997}.

A new class of states of the quantized electromagnetic field exhibiting unequal quadrature noise, nowadays known as {\it squeezed states}, was envisaged by E.H. Kennard in 1927 \cite{kennard1927} and later rediscovered and formalized by D. Stoler \cite{stoler1970} and H.P. Yuen \cite{yuen1976}. The {\it ideal squeezed states} are minimum uncertainty states that saturate the Heisenberg bound, i.e., $\Delta X \Delta Y = 1/4$.  In this case, we may write $\Delta X = e^{-r}/2$ and $\Delta Y = e^{r}/2$, where $r \geq 0$ is the modulus of the complex {\it squeezing parameter} $\zeta = r e^{i\theta}$ \cite{loudonknight1987}. The squeezed states were first generated in the laboratory in 1985 \cite{slusher1985}. Their peculiar quantum statistical properties make them a valuable resource for protocols that rely on quantum noise reduction and manipulation, including continuous-variable quantum cryptography \cite{hillery2000,oruganti2025} and quantum-enhanced metrology \cite{caves1981,desouza2014,desouza2021}, as demonstrated in gravitational-wave detection with LIGO \cite{LIGO2019}. Within the open quantum systems framework, squeezed states can be used to engineer structured environments. A system coupled to a broadband squeezed field exhibits phase-sensitive dissipation and altered relaxation dynamics \cite{gardiner1986}, and even when the environment is reduced to a single squeezed mode, it can induce substantial changes in the evolution of the system  \cite{decordi2023}. Gaussian states are quantum states of light that are entirely determined by their first and second quadrature moments, with all higher-order cumulants vanishing, or equivalently, quantum states whose Wigner functions are Gaussian functions in phase space. The most general pure single-mode Gaussian states are the squeezed coherent states, which can be defined as eigenstates of the Bogoliubov-transformed annihilation operator, $(\mu \hat{a} + \nu \hat{a}^\dagger) |\alpha,\zeta\rangle = \lambda |\alpha,\zeta\rangle$, with $|\mu|^2 -|\nu|^2 = 1$ plus the relations $\mu = \cosh r$, $\nu = e^{i\theta} \sinh r$, and $\lambda = \mu\alpha + \nu\alpha^*$ \cite{yuen1976}. Squeezed coherent states can exhibit photon-number fluctuations either larger or smaller than those of coherent states, corresponding to super-Poissonian or sub-Poissonian statistics, depending on the parameters $\alpha$ and $\zeta$ \cite{loudonknight1987}. In particular, the squeezed vacuum state ($\alpha = 0$ and $r\neq0$) is always super-Poissonian.

Despite the undeniable usefulness of Gaussian states, several quantum technologies require states having non-Gaussian elements. Beyond Gaussian limits, one can theoretically define and experimentally realize various non-Gaussian states \cite{walschaers2021}. The canonical example of a non-Gaussian, nonclassical state is the Fock state, which exhibits sub-Poissonian statistics at the lower bound due to its vanishing photon-number variance. The manipulation of quantum states of light in the Fock basis offers a natural and versatile route to the generation of non-Gaussian states. Among the possible methods to accomplish this we may cite the {\it quantum state truncation} or {\it quantum scissors} \cite{barnett1998,mattos21} and {\it Fock state filtering} processes, the latter also known as {\it hole burning} \cite{baseia1998,baseia2004,zeilinger2006,sivakumar2018,mattos2023}. 

Consider a generic state of light expressed in the Fock state basis $|\psi\rangle = \sum A_n |n\rangle$. If the $m$-th Fock component is removed, we end up with the following Fock filtered state 
\begin{equation}
	|\psi'\rangle = N_m\sum_{\substack{n = 0 \\ n \neq m}}^\infty A_n |n\rangle,
\end{equation}
where $N_m$ is a normalization constant. Note that this process is fundamentally different from the {\it photon subtraction} process, that is, the application of the annihilation operator to a quantum state of light, $\hat{a}|\psi\rangle$ \cite{agarwal07}. Photon subtraction is another viable mechanism for generating nonclassical states of light and can be applied to any state $|\psi\rangle$ other than a coherent state.

In this work, we introduce the Fock-state-filtered squeezed coherent states of a single-mode cavity field, a class of non-Gaussian states that have a photon number distribution very close to that of the squeezed coherent states, except for a ``hole" in the $m$-th Fock component. We investigate the dependence of photon-number statistics on the filtering index $m$, as well as on the displacement amplitude $\alpha$ and the squeezing parameter $\zeta$ that define the original squeezed state. We find that the filtering process significantly impacts the statistical properties of the state, and that nonclassical features such as sub-Poissonian behaviour can be enhanced. Additionally, the field may exhibit stronger super-Poissonian statistics than the original coherent state. Furthermore, we analyze how the nonclassical features of the states evolve when the system is subjected to a thermal environment. As expected, states exhibiting enhanced nonclassicality are more susceptible to environmental degradation. Nevertheless, within the parameter regime investigated here, the advantage in photon-number statistics introduced by the filtering process is not immediately erased and persists during part of the evolution toward the thermal equilibrium state.  

Our paper is organized as follows: in Section (\ref{sec2}), we introduce Fock-state-filtered squeezed coherent states and analyze their photon-number fluctuations, as quantified by Mandel’s $Q$ parameter, focusing on states with the Fock components either the $m = 0$ or the $m = 1$ components removed. We also discuss the nonclassicality of the filtered states using the Klyshko criterion \cite{klyshko1996}. In Section (\ref{sec3}), we examine the effects of coupling between the cavity field and a thermal bath. In Section (\ref{sec4}), we present our conclusions.

\section{Fock-state-filtered squeezed coherent states}\label{sec2}

\subsection{Definition and generation}

The squeezed coherent states can be defined as the states resulting from the successive application of the squeezing operator, $\hat{S}\left(\zeta\right)=\exp\left[\frac{1}{2}\left(\zeta^{*}\hat{a}^{2}-\,\zeta\,\hat{a}^{\dagger2}\right)\right]$ and the displacement operator, $\hat{D}\left(\hat{\alpha}\right)=\exp(\,\alpha\,\hat{a}^{\dagger}-\,\alpha^{*}\hat{a})$ to the vacuum state $|0\rangle$:
\begin{equation}
	\left|\alpha,\,\zeta\right\rangle = \hat{D}\left(\alpha\right)\hat{S}\left(\zeta\right)\left|0\right\rangle,     
\end{equation}
where we write $\alpha = |\alpha|e^{i\phi}$ and $\zeta = r e^{i\theta}$. The coefficients of the corresponding Fock state basis expansion, $|\alpha,\zeta\rangle = \sum_{n = 0}^\infty C_n |n\rangle$ are given by \cite{loudonknight1987}
\begin{equation}
	C_n = \frac{1}{\sqrt{n!\cosh r}} \left(\frac{1}{2} e^{i\theta}\tanh r\right)^{n/2}\exp\!\left[-\frac{1}{2}\left(|\alpha|^2+\alpha^{*2} e^{i\theta}\tanh r\right)\right] H_n\!\left( \frac{\alpha+\alpha^* e^{i\theta}\tanh r}{\sqrt{2 e^{i\theta}\tanh r}}\right),\label{coeffssqueezed}
\end{equation}
in terms of $H_n$ the Hermite polynomial of degree $n$. Thus, the photon-number distribution of squeezed coherent states, $P_n = |C_n|^2$, has an involved analytical form and may exhibit pronounced oscillations \cite{schleich87}.

We define the Fock-state filtered squeezed coherent states as 
\begin{equation}
	|\alpha,\zeta;m\rangle = N_m \sum_{\substack{n = 0 \\ n \neq m}}^\infty C_n |n\rangle,\label{fockfilteredstate}
\end{equation}
where $N_m = (1 - |C_m|^2)^{-1/2}$ is the normalization constant. 

The generation of the states in Eq.(\ref{fockfilteredstate}) can be accomplished in a cavity QED scheme, as the one proposed in \cite{liu2018}, where a two-level atom (frequency $\omega$ and states $|g\rangle$ and $|e\rangle$) interacts, with coupling constant $\lambda$, with an initial cavity quantized field $|\psi\rangle=\sum_{n=0}^{\infty} C_n |n\rangle$ (frequency $\omega_c$). If the field is highly detuned from the atom, $\Delta=\omega-\omega_c \gg \lambda\sqrt{n}$ (dispersive approximation), the atom-field dynamics will be described by the following effective Hamiltonian derived within the Jaynes-Cummings model framework,
\begin{equation}
	H_{\rm eff}=\frac{\Delta}{2}\left(1+\frac{2\lambda^2}{\Delta^2}a^\dagger a\right)\sigma_z,
\end{equation}
which produces photon-number-dependent shifts of the atomic transition frequency. The transition associated with the Fock component $|m\rangle$ can be selectively addressed by a classical field with strength $\Omega_m$ and photon-number–dependent, tunable frequency
\begin{equation}
	\omega_m = \omega + (2m + 1)\frac{\lambda^2}{\Delta}.
\end{equation}
Under the condition $\Omega_m \ll \lambda^2/\Delta$, only the component $|g,m\rangle$ undergoes a resonant Rabi oscillation,
\begin{equation}
	|g,m\rangle \rightarrow \cos(\Omega_m t)|g,m\rangle -i\sin(\Omega_m t)|e,m\rangle,
\end{equation}
while all other Fock components remain unaffected. After a half Rabi cycle, $t = \pi/2\Omega_m$, the resulting state will be
\begin{equation}
	|\Psi\rangle \propto \sum_{n\neq m} C_n |g,n\rangle -i C_m |e,m\rangle.
\end{equation}
A projective measurement leaving the atom in the ground state $|g\rangle$ then collapses the cavity field into the filtered state in Eq.(\ref{fockfilteredstate}). We should remark that the generation of filtered squeezed coherent states poses further challenges when compared to the filtering of coherent states. First, the preparation of squeezed states requires additional experimental resources. The generation of squeezed states generally depends on nonlinear optical processes such as four-wave mixing \cite{slusher1985}, but we may also cite a cavity QED scheme in which a three-level atom undergoes dispersive interactions while also being driven by a classical field \cite{moussa03}. A possible experimental method of generating of filtered squeezed states could be based on the squeezed-state preparation protocol of Ref. \cite{moussa03} and the filtering scheme described above \cite{liu2018}, within the same cavity, according to the following sequence: i) inject a coherent state in the cavity containing a three-level atom; ii) apply external driving fields to engineer the effective squeezing Hamiltonian and generate the squeezed state; iii) reconfigure the setup (driving fields, adjust frequencies, etc.); iv) start the filtering protocol by injecting a new atom in the cavity; v) run the filtering protocol. Of course, such a scheme demands precise control of the atom-field interaction and system stability characteristic of cavity QED experiments. Regarding the Fock state filtering part of the protocol, the atom-cavity detuning must be large compared with the coupling strength, $\Delta \gg \lambda \sqrt{n}$, while at the same time remaining small enough to satisfy $\Delta \ll \lambda^2/\Omega_m$, which requires high control and a careful choice of parameters. Also, compared with coherent states, squeezed coherent states, being intrinsically nonclassical, are generally more susceptible to cavity losses and other decoherence mechanisms, reflecting the well-known fragility of nonclassical states under environmental interactions \cite{davidovich96,souza08,barbosa10}. Nevertheless, we expect that the protocol should be experimentally feasible while keeping most of the nonclassical features of the seed state, provided that the interaction times remain short compared with the relevant decoherence timescales. In particular, cavity losses, occurring at a rate $\kappa$ require the generation time to satisfy $t_g = \pi/{2\Omega_m} \ll 1/\kappa$, thus minimizing photon leakage during the process. Nevertheless, any loss of the cavity photons will result in a ``refilling" of the holes, degrading the filtering process. For instance, having a finite probability for the loss of a single photon from the $|1\rangle$ component leads to $|1\rangle \rightarrow |0\rangle$, thus recovering the vacuum state component. Regarding the atomic spontaneous emission (at a rate $\gamma$), since the filtering protocol relies on the post-selection of the atom in the ground state $|g\rangle$ after the atom-field coherent evolution has taken place, atomic decay occurring before the measurement will be detrimental to the process. In dispersive cavity-QED implementations employing long-lived atomic states, cavity losses are usually the dominant source of decoherence, while atomic spontaneous emission is typically less important. Nevertheless, it is essential to keep the cavity and atomic decay rates as low as possible, so that the generation time remains much shorter than the typical decay times, $t_g \ll 1/\kappa, 1/\gamma$.

\subsection{Nonclassicality of the filtered states}

Squeezed states are Gaussian, nonclassical states characterized by a reduction of noise in one quadrature below the vacuum level, while they may also exhibit either sub-Poissonian or super-Poissonian photon statistics. The removal of a single Fock component is likely to have an impact on the nonclassical properties of the resulting Fock-filtered state. While we do not expect a reduction of quadrature noise compared to the original state, assumed as being a minimum-uncertainty squeezed coherent state, the filtering process directly modulates the photon number distribution of the state, and we thus expect changes in its photon statistics. Specifically, we are going to seek whether it is possible to amplify the sub-Poissonian character of a state that is already sub-Poissonian. 

To characterize the photon statistics, we calculate Mandel's $Q$ parameter, defined as
\begin{equation}
	Q = \frac{\langle (\Delta \hat n)^2 \rangle - \langle \hat n \rangle} {\langle \hat n \rangle}.\label{mandelq}
\end{equation}
This parameter quantifies the deviation from the Poissonian statistics of a coherent state, for which $Q = 0$. If $Q > 0$, the quantum state of light has super-Poissonian statistics while if $Q < 0$ the state is sub-Poissonian.
For the squeezed coherent state $|\alpha,\zeta\rangle$, with $\alpha  = |\alpha| e^{i\phi}$ and $\zeta = r e^{i\theta}$, Mandel's $Q$ parameter reads
\begin{equation}
	Q_{sq} = \frac{|\alpha|^2 \left[ \cosh(2r) - \sinh(2r) \cos(2\phi - \theta) \right] + \frac{1}{2} \sinh^2(2r)}{|\alpha|^2 + \sinh^2 r} - 1.\label{mandelqsq}
\end{equation}

The squeezed vacuum state $|0,\zeta\rangle = \hat{S}(\zeta) |0\rangle$ has $Q_{sv} = 2\sinh{r^2} + 1$, i.e., is always super-Poissonian \cite{loudonknight1987}. Its expansion in the Fock state basis is given by 
\begin{equation}
	|0,\zeta\rangle = \frac{1}{\sqrt{\cosh r}} \sum_{n=0}^{\infty} \left(-e^{i\theta}\tanh r\right)^n\frac{\sqrt{(2n)!}}{2^n n!}\,|2n\rangle.
\end{equation}
This corresponds to a broad photon number distribution which is restricted to even photon numbers, reflecting the pairwise generation of photons and the resulting photon bunching.

For squeezed coherent states, the photon-number fluctuations exhibit a richer structure, as they result from the combined effects of coherent displacement and quadrature squeezing. For small $|\alpha|^2$, being close to the squeezed vacuum, the state remains super-Poissonian. However, as the coherent amplitude increases, the state may become sub-Poissonian depending on the parameters. For a large coherent amplitude, $|\alpha|^2 \gg \sinh^2{r}$, and considering $\phi = \theta = 0$, we obtain from Eq.(\ref{mandelqsq}),
\begin{equation}
	Q_{sq} \approx e^{-2 r} - 1 < 0,
\end{equation}
corresponding to sub-Poissonian statistics. In this case, the displacement lies along the squeezed quadrature direction, which may lead to reduced photon number fluctuations, i.e., sub-Poissonian character, whereas if the displacement lies along the anti-squeezed quadrature direction, the statistics of the field will be super-Poissonian.

We now consider the states generated by filtering of specific Fock components of the squeezed coherent state, and investigate whether this process can enhance the nonclassical features of the original squeezed coherent state. The Mandel $Q$ parameter of the Fock-state-filtered squeezed coherent state having $m$ Fock component removed is:
\begin{equation}
	\begin{split}
		Q_{f} &= \frac{|\alpha|^2 [\cosh(2r) - \sinh(2r) \cos(2\phi - \theta)] + \frac{1}{2} \sinh^2(2r)}{(1 - P_m) \left( |\alpha|^2 + \sinh^2 r - m P_m \right)} \\
		&\quad - \frac{P_m \left\{ |\alpha|^2 [\cosh(2r) - \sinh(2r) \cos(2\phi - \theta)] + \frac{1}{2} \sinh^2(2r) + \left( |\alpha|^2 + \sinh^2 r - m \right)^2 \right\}}{(1 - P_m) \left( |\alpha|^2 + \sinh^2 r - m P_m \right)} - 1,
	\end{split}
\end{equation}
Here $P_m = |C_m|^2$, with $C_m$ defined in Eq.(\ref{coeffssqueezed}). For simplicity, we restrict our analysis to initial squeezed states with real positive parameters $\alpha$ and $r$ ($\phi = \theta = 0$). To illustrate the effects of the filtering process, we will consider two representative cases: the removal of either the $m = 0$ or the $m = 1$ Fock components from a squeezed coherent state having $r = 0.412$. This is the value of $r$, along with $\alpha = 1.64$, for which the Mandel $Q$ parameter reaches its minimum value within the ranges $0 \leq \alpha \leq 2.0$ and $0 \leq r \leq 1.2$ in a filtered state with the component $m = 1$ removed. After fixing $r$, we vary the coherent amplitude $\alpha \ge 0$, which is used here as a control parameter in our analysis. We focused on the filtering of either the $m = 0$ or the $m = 1$ components for states with a relatively low mean photon number because this favors both an enhancement of nonclassical effects and the experimental feasibility of the protocol. Removing higher-$m$ components from weak fields would have little impact on the resulting state due to the small occupation probabilities $P_m$. Besides, for more intense fields, where higher-$m$ components acquire significant population, the filtering process becomes increasingly challenging, since the dispersive approximation, $\Delta=\omega-\omega_c \gg \lambda\sqrt{n}$, must remain valid over the relevant photon-number distribution. This condition is more easily fulfilled when the field is concentrated in low-photon-number states. Thus, the choice of $m = 0,1$ provides a compromise between enhancing nonclassical effects and preserving the validity of the dispersive approximation.

\subsubsection{Filtering of the vacuum component ($m=0$)}

We first analyze the case $m=0$, which corresponds to the removal of the vacuum component from a squeezed state. This can significantly modify the photon-number statistics when the mean photon number is not large and lead to nonclassicality. In Fig.(\ref{figure1}a) we have plots of Mandel's $Q$ parameter of a squeezed coherent state and a Fock filtered state with $m = 0$ for $r = 0.412$ as a function of $\alpha$. The squeezed coherent state is super-Poissonian for small $\alpha$, and as the coherent amplitude becomes larger, we note a transition to sub-Poissonian statistics, reaching, in that interval, the minimum value of $Q_{sq} \approx -0.49$. In contrast, the filtered state is sub-Poissonian, reaching values below the minimum obtained for the squeezed coherent state within this interval. We should highlight the particular case of taking the limit of small $\alpha$ and $r = 0.0$, which, according to Eq.(\ref{mandelq}), results in $Q_f \approx -1$, practically the minimum possible value for Mandel's $Q$ parameter. This is because, by removing the vacuum component ($m = 0$) and having $\alpha$ small, the state effectively becomes the single photon state, i.e., $|\alpha,\zeta=0;m=0\rangle \approx \alpha |1\rangle$ (before normalization).  

\subsubsection{Filtering of the single-photon component ($m=1$)}

\begin{figure}[htbp]
	\centering
	\begin{subfigure}[b]{0.48\textwidth}
		\centering
		\includegraphics[width=\textwidth]{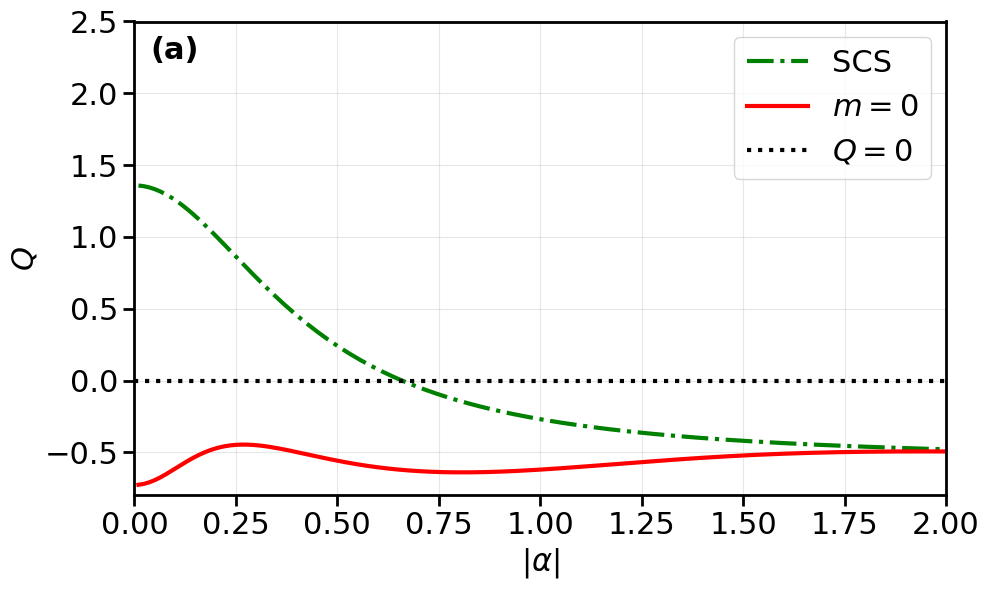}
		\label{figure1a}
	\end{subfigure}
	\hfill 
	\begin{subfigure}[b]{0.48\textwidth}
		\centering
		\includegraphics[width=\textwidth]{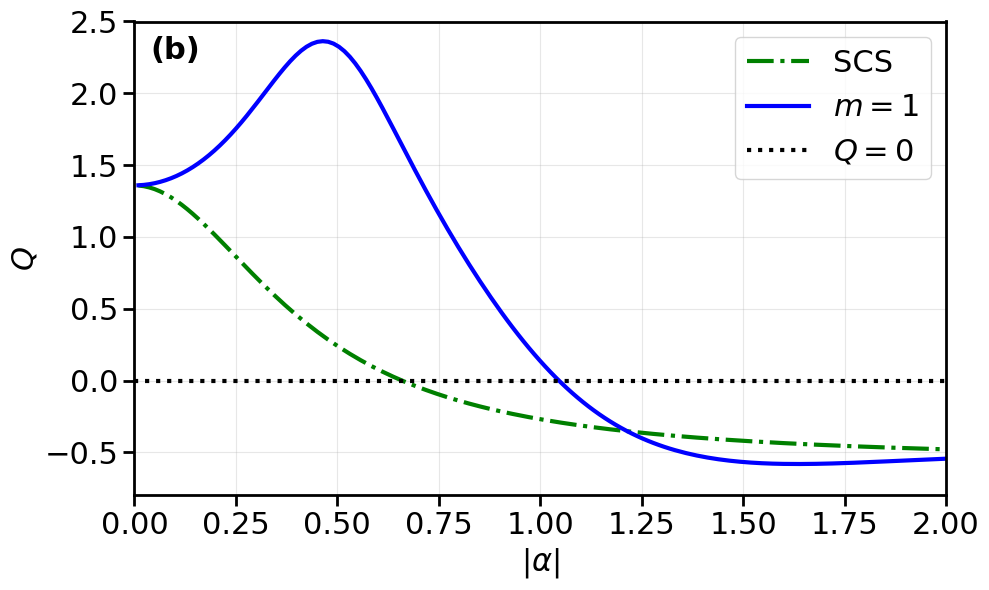}
		\label{figure1b}
	\end{subfigure}
	\caption{Mandel's $Q$ parameter as a function of the coherent displacement $\alpha$ (control parameter) for a squeezed coherent state with $r=0.412$ (green dot-dashed line) and Fock-filtered states with: (a) $m = 0$ (red continuous line) and (b) $m = 1$ (blue continuous line) components removed.}
	\label{figure1}
\end{figure}

We now consider the case $m=1$, which corresponds to the removal of the single-photon component from a squeezed state. Unlike the vacuum state filtering case, the effect of this operation on the photon statistics is less straightforward and may lead to qualitatively different behaviour. In fact, as shown in Fig.(\ref{figure1}b), for $\alpha \rightarrow 0$ the parameter $Q$ is that of the squeezed vacuum state for both the original squeezed state and the filtered state. Note that the $m = 1$ Fock component has been removed, but this has no effect, since the squeezed vacuum has only even Fock components. As $\alpha$ increases, the state initially exhibits enhanced super-Poissonian character. This rise in the $Q$ parameter is due to the increased weight of the $|0\rangle$ and $|2\rangle$ components, which, in the absence of the $|1\rangle$ term, effectively broadens the photon number distribution. We can also compare this effect with typical photon number fluctuations in a thermal state having the same mean photon number. The mean photon number of the Fock-filtered squeezed coherent state is
\begin{equation}
	\langle\hat{n}\rangle_f = \frac{|\alpha|^2 + \sinh^2 r - m\,|C_m|^2}{1-|C_m|^2},\label{nmediofiltered}
\end{equation}
which gives $\langle\hat{n}\rangle_f \approx 0.146$ when $Q_f^{(max)}\approx 2.36$, compared to $Q_{th} =  \langle\hat{n}\rangle_{th} =  0.146$ for the thermal state. An even more pronounced effect associated with photon number fluctuations, is the appearance of {\it super-bunching} in the Fock-filtered state, as quantified by the second-order normalized correlation function, 
\begin{equation}
	g^{(2)}(0) = 1 + \frac{Q}{\langle\hat{n}\rangle}.
\end{equation} 
While any thermal state has a maximum value of $g^{(2)}_{th}(0) = 2$, the filtered state with  $Q_f^{(max)} = 2.36$ has $g_f^{(2)}(0) \approx 17$, indicating the super-bunching caused by the removal of the $m  =1$ Fock component from the original squeezed state for small enough $\alpha$. We note that the original squeezed state can also be bunched, although to a lesser extent. For instance, if we take $\langle\hat{n}\rangle \approx 0.397$, that is, the value of the mean photon number corresponding to the peak of $Q$ in the filtered state, we have that the second order correlation function for a squeezed coherent state with the same $\langle\hat{n}\rangle$ is $g^{(2)}_{sq}(0) \approx 1.78$. However, we should shift the filtered state away from its noise peak to a point corresponding to the mean photon number $\langle\hat{n}\rangle_f \approx 0.397$ resulting in $Q_f \approx 1.45$ and $g_f^{(2)}(0) \approx 4.65$ for the filtered state. Thus, in this example, the Fock-state filtered squeezed coherent state is approximately $2.6$ times more bunched than a squeezed coherent state having the same mean photon number.

Further increasing $\alpha$ moves the photon number distribution to higher photon numbers; Mandel's $Q$ parameter decreases, and the filtered state becomes sub-Poissonian. Interestingly, $Q$ reaches a minimum value of approximately $Q_f \approx -0.58$, which is lower than the minimum value obtained for the squeezed coherent state over the same range of $\alpha$, as seen in Fig.(\ref{figure1}). This occurs because, as $\alpha$ increases and the photon-number distribution shifts toward higher $n$, removing the $m=1$ component effectively suppresses the low-photon-number part of the distribution, leading to a narrower photon-number spread and an enhanced sub-Poissonian character. Such a narrowing is expected to be more pronounced in the present case, where a relatively weak cavity field is considered.

We remark that although the quadrature squeezing is not expected to increase, given that the original state is already a minimum-uncertainty squeezed state, a scan of the parameters $\alpha$ and $r$ of the filtered state showed that it is possible to have reduction in the photon number fluctuations in relation to the original squeezed state while still exhibiting quadrature squeezing.

\subsubsection{Klyshko criterion}

The analysis done via the Mandel $Q$ parameter can be complemented using the Klyshko criterion \cite{klyshko1996}, an operational witness of nonclassicality which compares neighboring values of probabilities of the field's photon number distribution, e.g., $P_n$, $P_{n+1}$, and $P_{n+2}$. A simple sufficient criterion for the nonclassicality of a light field is provided by the Klyshko parameter $B(n)$, defined as
\begin{equation}
	B(n)=(n+2)P_nP_{n+2}-(n+1)P_{n+1}^2.
\end{equation}
If
\begin{equation}
	B(n)<0
\end{equation}
for at least one value $n$, the state is necessarily nonclassical. On the other hand, if $B(n) \geq 0$, the test is inconclusive, since the state may still have nonclassical features not detected by this witness. The Klyshko criterion is closely related to the Glauber-Sudarshan $P$-function \cite{sudarshan1963,glauber1963}; when $B(n) < 0$, the photon number distribution of the field exhibits a local structure not possible in ``classical" fields. For instance, in the case of a coherent state, $B(n) = 0$ identically for all $n$.

The Klyshko parameter for the Fock filtered squeezed coherent states reads
\begin{equation}
	B_f(n) = \frac{(n+2)P_nP_{n+2}\left(1-\delta_{n,m}\right)\left(1-\delta_{n+2,m}\right)-(n+1)P_{n+1}^{\,2}\left(1-\delta_{n+1,m}\right)}{\left(1-P_m\right)^2}.\label{Klyshkofiltered}
\end{equation}
Thus, for $n = m$,
\begin{equation}
	B_f(n) = -\frac{(m+1)P_{m+1}^{\,2}}{\left(1-P_m\right)^2} < 0,
\end{equation}
meaning that the filtered states here considered are always nonclassical according to this criterion.

\section{Effects of a thermal bath on nonclassicality}\label{sec3}

We now investigate how the nonclassical properties of the Fock-filtered states are affected by dissipation. We consider the coupling of the cavity field to a thermal bath and describe the dynamics using the standard Markovian master equation for a damped cavity in the Lindblad form. The master equation in Eq.(\ref{mastereqgen}), for the reduced system's density operator $\rho_s$, written in the interaction representation for this scenario, is given by \cite{breuer02}
\begin{equation}
	\frac{d\rho_s}{dt} = \gamma(\bar{n}+1)\left(2\hat{a}\rho_s\hat{a}^\dagger - \hat{a}^\dagger \hat{a}\rho_s - \rho_s \hat{a}^\dagger \hat{a}\right) + \gamma \bar{n}\left(2\hat{a}^\dagger \rho_s\hat{a}-\hat{a}\hat{a}^\dagger \rho_s - \rho_s \hat{a}\hat{a}^\dagger\right),\label{mastereqbath}
\end{equation}
where 
\begin{equation}
	\overline{n}=\frac{1}{e^{\hbar \omega/k_B T}-1}
\end{equation}
is the mean photon number of the thermal state of the bath at a temperature $T$ and $\gamma$ is the decay constant, related to the coupling of the cavity field mode of frequency $\omega$ to the thermal bath. If $T = 0$ K, for which $\overline{n} = 0$, the cavity field undergoes pure damping and loses energy to the environment, as described by the remaining term on the right-hand side of Eq.(\ref{mastereqbath}). However, for $T \neq 0$ K, we have both photon loss and absorption processes, so that thermal photons from the bath can also excite the cavity field mode. The following results were obtained via the numerical solution of Eq.(\ref{mastereqbath}).

\begin{figure}[htbp]
	\centering
	\begin{subfigure}[b]{0.48\textwidth}
		\centering
		\includegraphics[width=\textwidth]{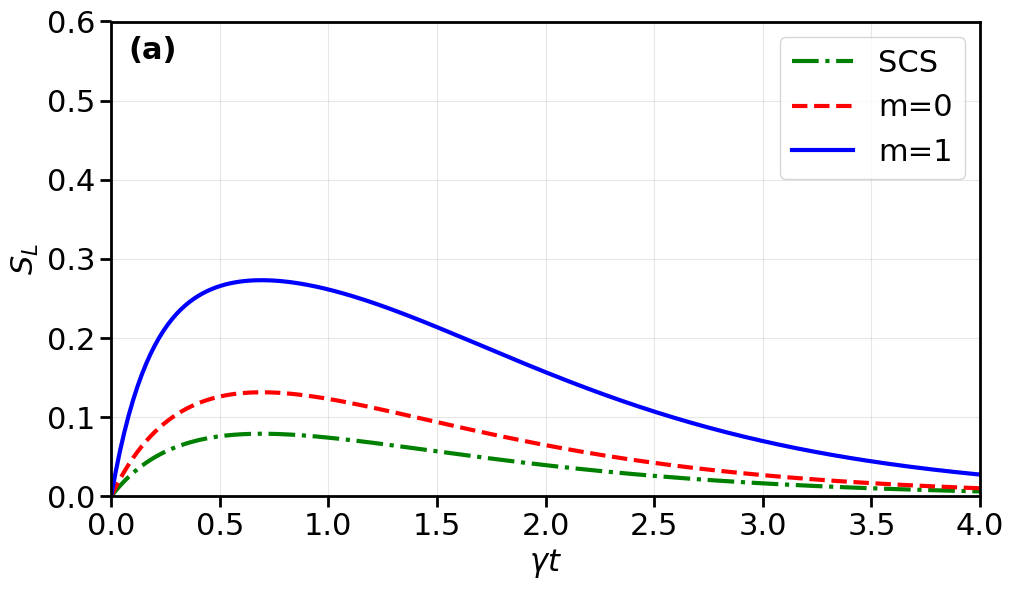}
		\label{figure2a}
	\end{subfigure}
	\hfill 
	\begin{subfigure}[b]{0.48\textwidth}
		\centering
		\includegraphics[width=\textwidth]{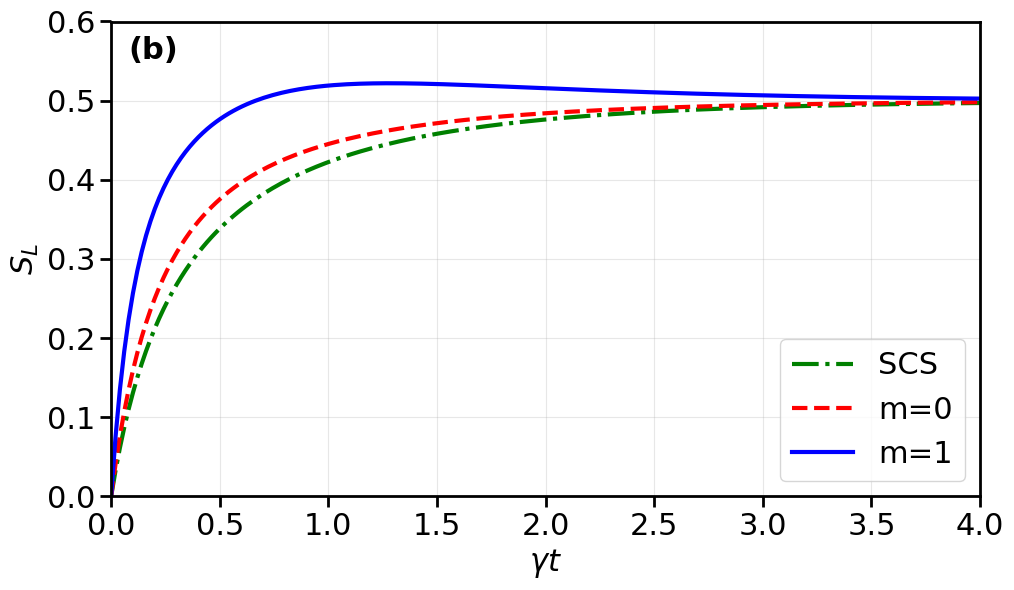}
		\label{figure2b}
	\end{subfigure}
	\caption{Linear entropy $S_L$ of the cavity field state, as a function of the normalized time $\gamma t$, for a field coupled to a thermal bath. The initial states are a squeezed coherent state with $r=0.412$ and $\alpha = 1.64$ (green dot-dashed line) and Fock-filtered states with either the $m = 0$ (red dashed line) or the $m = 1$ (blue continuous line) component removed; (a) zero temperature bath, $\overline{n} = 0$; (b) finite temperature bath with $\overline{n} = 0.5$.}
	\label{figure2}
\end{figure}

While in general we expect a degradation of nonclassical features due to the interaction with an environment, this depends on the specific field state and the type of coupling between the field and the bath. For instance, considering a bath at $T = 0$ K and the master equation in Eq.(\ref{mastereqbath}), the coherent states are the only states that remain pure during the evolution \cite{dutra98}. However, if the bath is at $T \neq 0$ K, the initial coherent state evolves to a mixed state. We thus expect that different quantized field states will evolve distinctly when in contact with the a thermal bath. A convenient way to quantify the state purity of a quantum state during its evolution is via the linear entropy, defined as $S_L(t) = 1 - Tr\rho_s^2(t)$. We have that $S_L = 0$ for any pure quantum state and $S_L > 0$ for mixed states. In Fig.(\ref{figure2}) we have plots of $S_L$ as a function of $\gamma t$, for three initial field states: a squeezed coherent state and the Fock filtered states with either the $m = 0$ or the $m = 1$ components removed. In Fig.(\ref{figure2}a), the bath is assumed to be at zero temperature ($\overline{n} = 0$). We note that the loss of purity is more pronounced for the state having $m = 1$, and that the original squeezed coherent state is the most resistant to losses. For longer times, the purity goes to zero, as the cavity evolves toward the vacuum state, which is pure. If the bath is at finite temperature ($\overline{n} = 0.5$), purity is lost more rapidly, with the  $m=1$ filtered state exhibiting the fastest degradation of its purity, as shown in Fig.(\ref{figure2}b). In this regime, the incoming thermal noise drives the evolution of $S_L$ for different initial fields closer together, with all curves approaching the same thermal equilibrium value at long times. 

\begin{figure}[htbp]
	\centering
	\begin{subfigure}[b]{0.48\textwidth}
		\centering
		\includegraphics[width=\textwidth]{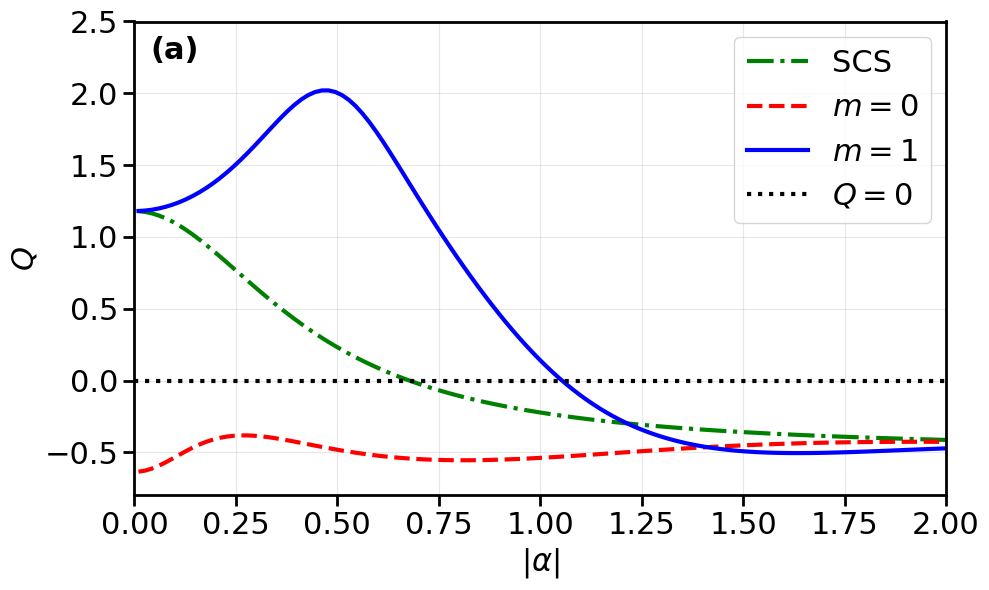}
		\label{figure3a}
	\end{subfigure}
	\hfill 
	\begin{subfigure}[b]{0.48\textwidth}
		\centering
		\includegraphics[width=\textwidth]{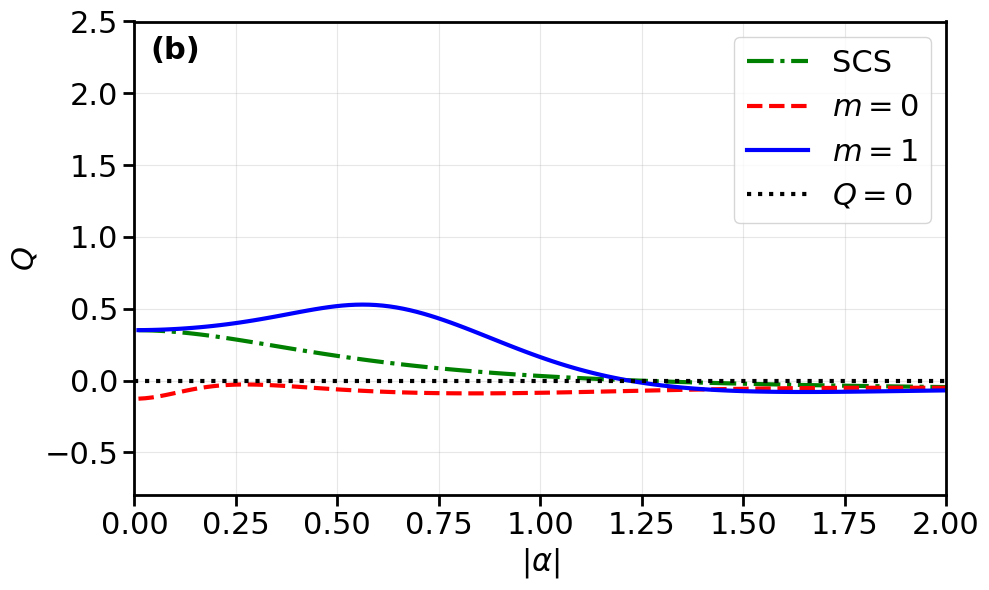}
		\label{figure3b}
	\end{subfigure}
	\caption{Mandel's $Q$ parameter as a function of the coherent displacement $\alpha$ (control parameter) for a squeezed coherent state with $r=0.412$ (green dot-dashed line), and Fock-filtered states with either the $m = 0$ (red dashed line) or the $m = 1$ (blue continuous line) components removed. The fields are in contact with a thermal bath ($\overline{n} = 0.1$) at two distinct normalized times: (a) $\gamma t = 0.1$; (b) $\gamma t = 1.0$.}
	\label{figure3}
\end{figure}

\begin{figure}[htbp]
	\centering
	\includegraphics[width=\textwidth]{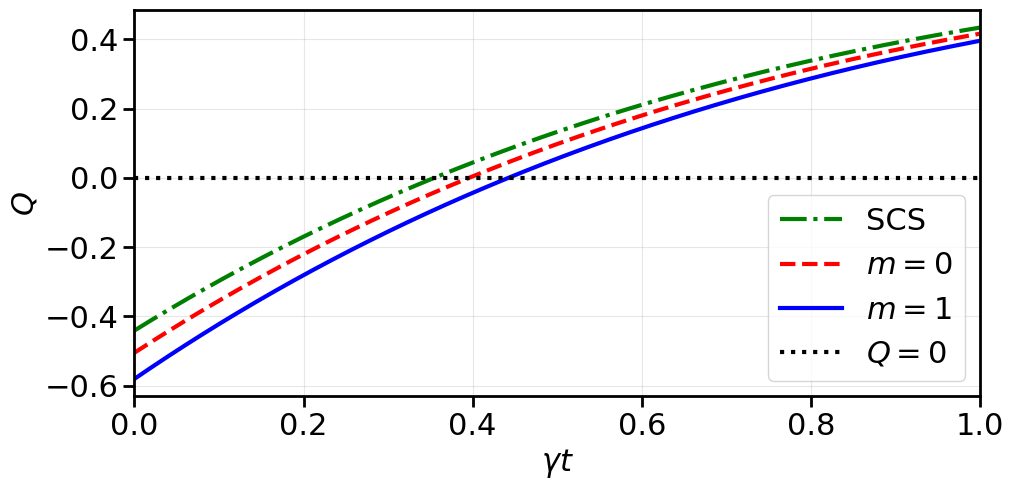}
	\caption{Mandel's $Q$ parameter as a function of the normalized time $\gamma t$, for a field coupled to a thermal bath with $\overline{n} = 0.1$. The initial states are a squeezed coherent state with $r=0.412$ and $\alpha = 1.64$ (green dot-dashed line) and Fock-filtered states with either the $m = 0$ (red dashed line) or the $m = 1$ (blue continuous line) component removed.}
	\label{figure4}
\end{figure}

The degradation of the nonclassical properties of the field is unavoidable if the system is in contact with a thermal bath. In Fig.(\ref{figure3}), we plot Mandel's $Q$ parameter as a function of $\alpha$ for two specific times during the evolution, $\gamma t = 0.1$ in Fig.(\ref{figure3}a), and $\gamma t = 1.0$ in Fig.(\ref{figure3}b). At the earlier time, the nonclassical features are still very prominent, while at $\gamma t = 1.0$ the curves are considerably more flattened. We note that, despite of the action of the environment causing decay of nonclassicality, the distinctive features of the filtered states, e.g., a $Q$ parameter smaller than that of the original squeezed state, are kept during the evolution, at least for the scenario considered here. From another perspective, we can fix the value of $\alpha = 1.64$, and calculate the time evolution of the $Q$ parameter. In Fig.(\ref{figure4}) we plot $Q$ as a function of $\gamma t$ for both Fock-filtered and squeezed coherent states, showing that Mandel's parameter for all three initial states evolves similarly toward the thermal, super-Poissonian equilibrium state. 

Gaussian states such as the original squeezed coherent state tend to be more robust under the action of the environment described by Eq.(\ref{mastereqbath}). In contrast, the interaction with the bath washes out the structure of the Fock-filtered states by incoherently redistributing photon-number populations and suppressing quantum coherences, which is reflected in the evolution of the linear entropy. On the other hand, Mandel's $Q$ parameter, which depends only on the lower moments of the distribution, $\langle \hat n \rangle$ and $\langle \hat n^2 \rangle$, has a smoother evolution and basically an exponential decay. As a consequence, under the specified conditions, the relative advantage provided by the filtering process is qualitatively maintained during part of the evolution, as shown in Fig.(\ref{figure4}).

\begin{figure}[htbp]
	\centering
	\begin{subfigure}[b]{0.48\textwidth}
		\centering
		\includegraphics[width=\textwidth]{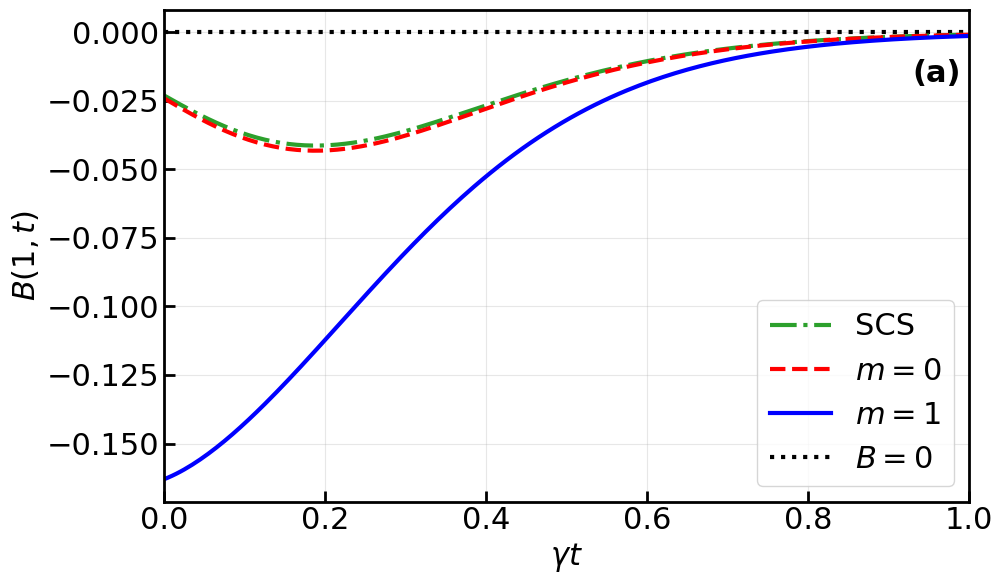}
		\label{figure5a}
	\end{subfigure}
	\hfill 
	\begin{subfigure}[b]{0.48\textwidth}
		\centering
		\includegraphics[width=\textwidth]{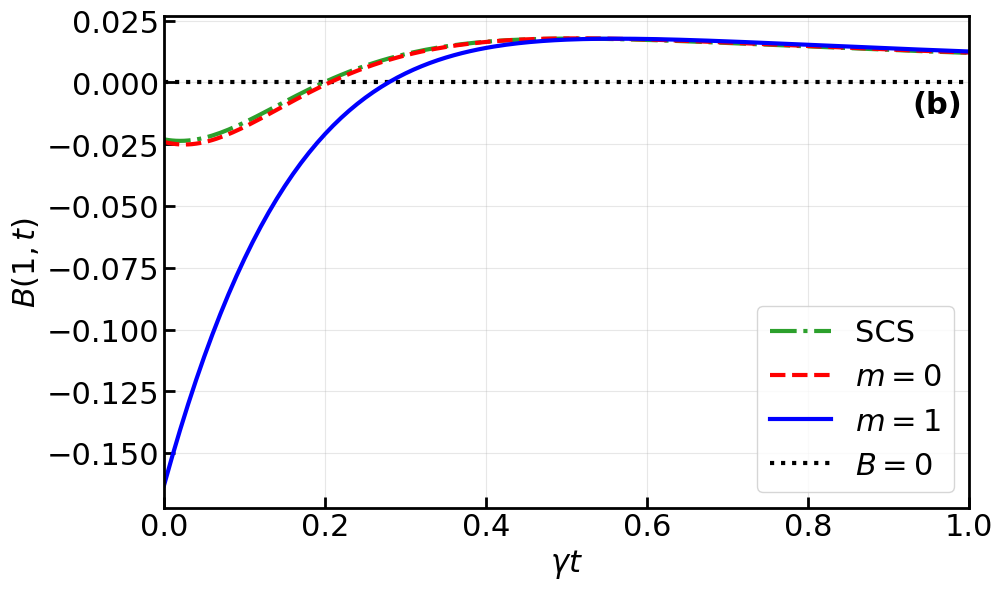}
		\label{figure5b}
	\end{subfigure}
	\caption{Klyshko parameter $B(1,t)$ of the cavity field state, as a function of the normalized time $\gamma t$, for a field coupled to a thermal bath. The initial states are a squeezed coherent state with $r=0.412$ and $\alpha = 1.64$ (green dot-dashed line) and Fock-filtered states with either the $m = 0$ (red dashed line) or the $m = 1$ (blue continuous line) component removed; (a) zero temperature bath, $\overline{n} = 0$; (b) finite temperature bath with $\overline{n} = 0.5$.}
	\label{figure5}
\end{figure}

We may also analyze the evolution of the Klyshko parameter, $B(n)$, during the cavity field dissipation process. In Fig.(\ref{figure5}), we plot $B(n,t)$ (for $n=1$) as a function of $\gamma t$, again for the three cases, the original squeezed coherent state, and the filtered states with either the $m = 0$ or the $m = 1$ components removed. We focus on the case $n=1$, which provides the clearest illustration of the effects of Fock-state filtering on the Klyshko criterion. We also investigated $B(n,t)$ for $n > 1$, and found that the corresponding curves exhibit the same qualitative behaviour during the dissipative evolution, but with weaker violations of the corresponding inequalities and shorter time intervals over which they certify nonclassicality. This is expected because $B(1,t)$ depends explicitly on the probabilities $P_1$, $P_2$, and $P_3$, making it particularly sensitive to the removal of the $|1\rangle$ Fock component. Consequently, $B(1,t)$ better characterizes the nonclassicality introduced by the $m = 1$ filtering process. As shown in Fig.(\ref{figure5}a), with the bath at $T = 0$ K, $B(1,t)$ approaches zero from below, and the $n = 1$ Klyshko inequality continues to witness nonclassicality over the displayed time interval, yet with decreasing strength. This is consistent with the fact that the cavity state at long times becomes the vacuum state $|0\rangle$. If the bath is thermally excited, e.g. with $\overline{n} = 0.5$, we observe the behaviour shown in Fig.(\ref{figure5}b): the curves cross $B(1,t) = 0$ at finite times and subsequently become positive as the field relaxes toward the classical thermal state. However, these crossings should not be identified as nonclassical-to-classical transitions, because $B(1,t) \geq 0$ only means that nonclassicality is no longer certified by $B(1)$. We remark that the information provided by the Klyshko parameter, shown in Fig.(\ref{figure5}), is complementary to that obtained from Mandel's $Q$ parameter. While the latter reveals actual transitions between sub-Poissonian and super-Poissonian photon statistics, as shown in Fig.(\ref{figure4}), the evolution of $B(1,t)$ identifies the time interval during which the Klyshko criterion certifies nonclassicality. We note that there is a stage of the field evolution in which the Klyshko parameter no longer provides such certification, even though the field remains sub-Poissonian and therefore nonclassical.

\section{Conclusion}\label{sec4}

In this work, we investigated the statistical properties and dissipative dynamics of Fock-filtered squeezed coherent states, with emphasis on photon-number fluctuations and nonclassicality as quantified by Mandel’s $Q$ parameter. We have shown that the selective removal of individual Fock components in a squeezed coherent state, the most general pure Gaussian state of light, may lead to a reshaping of the photon-number distribution, resulting in states with properties highly sensitive to the initial displacement $\alpha$. Depending on $\alpha$, used here as a control parameter, and the removed component $m$, the filtered states may exhibit enhanced sub-Poissonian behaviour or a pronounced super-bunching effect, compared to the original squeezed coherent states, as shown in Fig.(\ref{figure1}). The removal of low-photon-number components, e.g., $m = 1$ in the case of the relatively weak fields considered here, can suppress the low-photon-number sector, tending to narrow the photon-number distribution and thus enhancing the sub-Poissonian character of the field.

The influence of an environment was analyzed by assuming the coupling the cavity field to a thermal bath within the standard master-equation framework. As expected, all initial states evolve toward thermal statistics, leading to a monotonic decay of Mandel’s $Q$ parameter, which reflects its dependence on low-order moments of the photon-number distribution. For the cases investigated here, the statistical advantage introduced by the filtering process is qualitatively maintained during the field evolution. A complementary characterization is provided by the Klyshko criterion, which indicates how long the evolving field state remains certifiably nonclassical according to this criterion. We also found that the purity of the filtered states, quantified by the linear entropy $S_L$, is significantly more sensitive to the dissipative dynamics, which leads to a rapid and pronounced increase in mixedness even in the zero temperature limit, as shown in Fig.(\ref{figure2}). 

The filtered squeezed states constitute a new class of non-Gaussian continuous-variable states that simultaneously combine two important nonclassical resources: quadrature squeezing and reduced photon number fluctuations. Within the range of parameters investigated here, the filtering process typically reduces the degree of quadrature squeezing of the original state, while simultaneously enhancing its sub-Poissonian character. The simultaneous presence of these resources makes the filtered states interesting candidates for multi-parameter quantum estimation. In this context, sensing protocols involving the simultaneous estimation of an optical phase shift and an optical loss (attenuation) may benefit from probe states combining reduced quadrature and photon-number fluctuations. Besides, filtered states could also offer a versatile platform for exploring the interplay between distinct manifestations of nonclassicality and their evolution in realistic environments.

\medskip
\noindent
\textbf{Author contributions}
J.P.G.O. - calculations; numerical simulations; plots; formal analysis; review. A.V.-B. - conceptualization; writing; formal analysis; interpretation; review \& editing.

\medskip
\noindent
\textbf{Funding}
This work was partially supported by the Coordenação de Aperfeiçoamento de Pessoal de Nível Superior (CAPES), Brazil (Grant No. 88887.132329/2025-00).

\medskip
\noindent
\textbf{Data availability}
No datasets were generated or analysed during the current study.

\medskip

\section*{Declarations}
\noindent
\textbf{Conflict of interest}
The authors declare no conflict of interest.
\medskip

\noindent
\textbf{Competing interests}
The authors declare no competing interests.


\bibliography{refsfiltersq}
\bibliographystyle{unsrt}

\end{document}